\documentclass[aps,prl,twocolumn,groupedaddress]{revtex4}

\usepackage{graphicx}

\begin{document}


\title{Millisecond spin lifetimes in quantum dots at zero applied magnetic field due to strong electron-nuclear interaction}


\author{R. Oulton$^1$, A. Greilich$^1$, S. Yu. Verbin$^{1,\star}$, R.V. Cherbunin$^{1,\star}$, T. Auer$^1$, D.R.~Yakovlev$^1$, M. Bayer$^1$, V. Stavarache$^2$, D. Reuter$^2$, and A. Wieck$^2$}

\affiliation{$^1$Experimentelle Physik II, Universit\"at Dortmund,
D-44221 Dortmund, Germany} \affiliation{$^2$Angewandte
Festk\"orperphysik, Ruhr-Universit\"at Bochum, D-44780 Bochum,
Germany}

\date{\today}

\begin{abstract}
A key to achieving ultra-long electron spin memory in
quantum dots (QDs) at 0~$T$ is the polarization of the nuclei, such that the
electron spin is stabilized along the nuclear magnetic field.  We
demonstrate that spin-polarized electrons in n-doped QDs align the nuclear field via the hyperfine
interaction. A feedback onto the electrons occurs, leading
to stabilization of electron polarization. We suggest that the
coupled electron-nuclear system forms a rigid nuclear spin polaron state as predicted by I.A.~Merkulov \cite{Merkulov98}, for
which spin memory is retained over millisecond lifetimes.
\end{abstract}

\pacs{123456789}

\maketitle

New impetus for the use of QDs in quantum information processing
\cite{LossPRL00,ImamogluPRL99} has recently been gained by reports
of long electron ($e$) spin relaxation times: coherence times
$T_{2}$ of several $\mu$s \cite{Petta05, Colton04} and lifetimes $T_{1}$ of
several ms \cite{Kroutvar05} have been demonstrated. With
techniques \cite{Cortez02,DzhioevPSS99,Ikezawa04,Kalevich05,Kroutvar05} to charge QDs with
spin polarized $e$'s, it appears that the $e$ spin is a promising
candidate for a qu-bit. Such a stable $e$-spin results from the
suppression of spin relaxation mechanisms present in higher
dimensionality structures. An interaction that does not become
inefficient for QDs, however, and predicted to be the limiting
factor for $e$-spin orientation, is the hyperfine interaction with
the nuclei \cite{Merkulov02}. While only a partial dephasing of
the $e$ spins occurs over a $\mu$s timescale \cite{Braun05}, total
loss of $e$ spin alignment eventually occurs \cite{Merkulov02} due
to fluctuations of the randomly orientated nuclei.

\begin{figure}
\includegraphics[width=8cm]{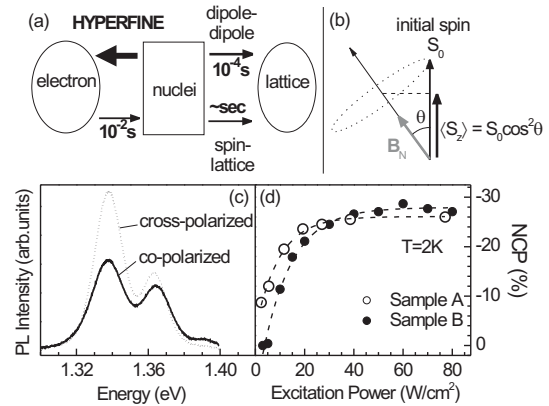}
\caption{(a) Interacting spin systems in QDs. (b) Sketch
of $e$ precession around nuclear field ${\bf B}_N$. (c)~Polarization-selective CCD PL spectra of QDs for
Sample A, with $\sigma^+$ polarized excitation. Detection was $\sigma^+$ or $\sigma^-$.
(d) NCP at peak PL intensity vs excitation density for Samples A (open circles) and B (closed circles). T~=~2K.} \label{Fig1}
\end{figure}

The spin dynamics in QDs are rather complex, as they are governed by
multiple interactions, as sketched in Fig.~\ref{Fig1}(a). Confined $e$'s interact with the
nuclei via the hyperfine interaction: An $e$ spin exerts a
magnetic field ${\bf b}_e$ (called Knight field) onto the nuclei
\cite{Meier}, which in combination act in turn on the $e$ (called Overhauser effect) to give
a magnetic field ${\bf B}_N = \sum_i^N {\bf
b}_{n,i}$, where the sum is over all $N$ nuclei in the QD.  This
hyperfine interaction has been shown to lead to transfer of
angular momentum from the $e$ to the nuclei via a spin "flip-flop", a process which occurs over a typical timescale of $10^{-2}$~s~\cite{Meier}. The magnitude of the total nuclear alignment
achieved, on the other hand, is influenced by the magnetic
dipole-dipole interaction between the nuclei, the
strongest depolarization process, acting on time scales of
$10^{-4}$ s. This interaction may be switched off by applying an external
magnetic field, such that the nuclear Zeeman splitting is larger
than the dipole-dipole interaction~\cite{Meier}. The nuclear interaction
with the lattice is the weakest of all, leading to a spin-lattice
relaxation time of seconds or longer, and is therefore
neglected in this paper.

Magnitude and direction of the nuclear field ${\bf B}_{N}$ depend
on the nuclear polarization degree. For typically $\sim 10^{5}$
nuclei in a self-assembled QD, $B_{N} \gg b_{e}$, and therefore the
nuclear precession frequency is orders of magnitude smaller than
that of the $e$, so that over a timescale of $\sim \mu s$ the $e$
experiences a constant, "frozen fluctuation" field ${\bf B}_N$
with a field strength of about 30 mT \cite{OultonNANO}. Due to
the $e$ precession, starting from an initial spin orientation
${\bf S}_{0}$, the spin polarization retained is the projection
$\langle S_z \rangle =  S_{0} cos^{2}\theta $ (see
Fig.~\ref{Fig1}b)~\cite{Merkulov02,Braun05,Imamoglu03}. An ensemble measurement probes a
distribution of $B_{N}$, and the $e$ spin-phase between QDs is
quickly lost, so that only an average spin $S_{0}/3$ is preserved~\cite{Merkulov02}. The key to increasing the spin polarization
therefore is to avoid this $e$ spin precession: either a strong
magnetic field needs to be applied \cite{Braun05} or the
nuclear spins should be aligned, such that $S_{0} cos^{2}\theta  = S_{0}$.

Nuclear
polarization may be achieved using the Overhauser effect
\cite{Meier,Paget77,Bracker05}, where the QD is constantly optically pumped
with spin polarized excitons. The nuclei experience the Knight
field from the $e$, and, as a result of a sequential multiple flip-flop processes, begin to align along the $e$ spin
polarization axis. Generally, an external magnetic field should be applied, leading to a splitting
of the nuclear spin levels above the energy for the nuclear
dipole-dipole interaction to occur. However, for QDs, the Knight
field $b_{e}$ at the site of a nucleus is comparable with the
field induced by a neighbouring nucleus \cite{Paget77}. Thus
$b_{e}$ alone may induce a nuclear Zeeman splitting that effectively
allows the hyperfine interaction to compete with the dipole-dipole
interaction, so that nuclear polarization may occur.\cite{Lai06} However, if
one introduces $e$'s by injecting neutral excitons, the Knight
field vanishes with the
radiative decay of the exciton.

Here we take a different, all-optical approach, by which
polarization of the $e$-nuclei system is achieved even without
external magnetic field.  A permanent field $b_e$ is maintained by
resident $e$'s in the QDs. These $e$'s are oriented by
circularly polarized laser excitation, for which we exploit an
optical pumping mechanism unique for n-doped QDs~\cite{DzhioevPSS99,Cortez02,Bracker05,Ikezawa04}.
The nuclei are polarized through the hyperfine interaction, and as a result of a
feedback by the nuclei, greatly stabilized $e$ polarization is
achieved.
Both spin systems are aligned such that the fluctuations of the
interacting $e$-nuclear system may be low enough to allow the
formation of a nuclear polaron state \cite{Merkulov98}: the spin
lifetime of this collective state is increased over orders of
magnitude to the millisecond scale.

The studied samples contained 20 layers of (In,Ga)As/GaAs
self-assembled QDs. A range of structures was addressed, results
from two of which are presented here: Samples A and B, thermally annealed
at 945 and 900$^{0}$C, with ground state emission energies at
$1.42$ and $1.34$ meV, respectively. The structures were n-doped 20
nm below each dot layer with a dopant density about equal to the
dot density, such that each dot was occupied with on average one
$e$. All measurements were performed at a temperature of $T$~=~2 K in an optical
cryostat, which was placed inside three pairs of Helmholtz coils,
oriented mutually orthogonal to each other. These coils were used
for compensating parasitic magnetic fields (e.g. the geomagnetic
field) down to $<1 ~\mu$T, and also to apply fields in the
Voigt/Faraday geometries. The QDs were excited with a
mode-locked Ti:Sp laser emitting pulses of width $\sim$ 1 ps
separated by $\sim$ 13 ns. The excitation beam was focused to a
spot size of $\sim 200 \mu$m diameter, the QD
photoluminescence (PL) dispersed by a monochromator and
resolved with a CCD camera, an avalanche
Si-photodiode (APD) or a streak camera with time resolution 30~ps.

Fig.~\ref{Fig1}(c) shows PL of the QDs after excitation with
$\sigma^+$ polarized light in the wetting layer at 1.476 eV,
detecting both co- ($\sigma^+\sigma^+$) and cross-polarized
($\sigma^+\sigma^-$) to the excitation with the CCD. The ground state PL
polarization, at maximum 1.34 eV, is negative, i.e. the QDs preferentially emit photons
with the opposite polarity to the exciting light. This negative
circular polarization (NCP) is now a well-established phenomenon in n-doped
QDs \cite{DzhioevPSS99,Cortez02,Bracker05,Ikezawa04}. The key feature
of this phenomenon is that the resident $e$'s become
polarized. While the polarization of these resident $e$'s in the ensemble may
be low after a single pulse, several successive pulses allow for
strong accumulation of the $e$ spin polarization, as reflected in the NCP value. Its dependence on
excitation power is given in Fig.~\ref{Fig1}(d): at low powers it increases, until saturation is reached at $\sim 20$~Wcm$^{-2}$.

Under these conditions, the system is prepared for polarization of
the nuclear system, for which we envisage the following scheme:
After recombination of the trion ($\sim 1$~ns), and before the next
laser pulse ($\sim 13$~ns), the QD contains a single polarized
$e$, which may transfer angular momentum in form of a flip-flop
process to the nucleus. At the same time the $e$ maintains an
internal magnetic field that produces a nuclear Zeeman splitting strong enough to suppress nuclear dipole-dipole dephasing. A
following laser pulse then reorients the $e$ spin, such that successive multiple pulses orient a significant number of nuclei and decrease the temperature of the nuclear system.

\begin{figure}
\includegraphics[width=8cm]{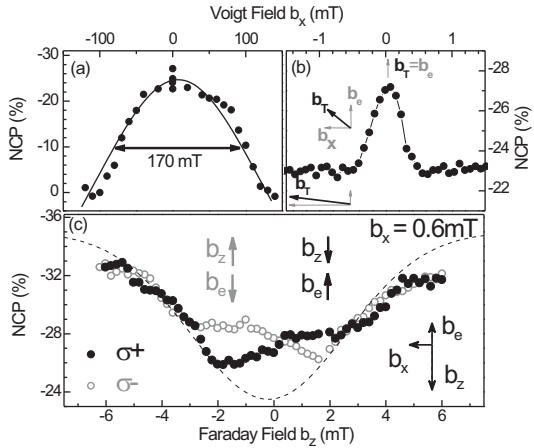}%
\caption{NCP for Sample B taken with APD at emission energy 1.34
eV as function of applied Voigt field $b_{x}$ over (a) 100 mT and
(b) 1 mT range. (c) NCP as function of Faraday field $b_z$ in
presence of Voigt field of 0.6 mT.  Lines in (a) and (c) are
guides to the eye.} \label{Fig2}
\end{figure}

The polarization of the nuclei leads to a total field ${\bf B}_N$ whose angle to the $z$-axis, $\theta $ is small. The NCP measured, determined by $\langle S_z
\rangle = S_{0} cos^{2}\theta $, should increase if there is a feedback onto the $e$.  We first
investigate whether the nuclei indeed affect the NCP, or whether
it can be explained solely by the spin-polarized excitation of the
$e$'s. The Knight field ${\bf b}_e$, can be estimated to be of the
order of 0.1 - 1 mT \cite{Paget77}. If relevant for creating a nuclear polarization, NCP
should be strongly affected by external fields of comparable
magnitude. Fig.~\ref{Fig2} (a) shows the influence of a much
stronger field (100 mT range) applied in Voigt geometry
perpendicular to the polarized $e$ spin. The NCP
reduces for field values $>80$ mT due to several processes, such as
the $e$ precession about the
external field (Hanle effect), and also the reduction of the anisotropic exchange \cite{DzhioevPSS99}. 

We turn now to the behavior of the NCP for very low fields $<2$mT.  As can be seen in Fig.~\ref{Fig2} the NCP value drops from a value of $\sim -28\%$ to a value of $\sim -23\%$ with a half-width at half maximum of just 0.25~mT.  Such low fields have a negligible direct effect on the carriers (for example, the Zeeman splitting induced is $\sim$10 neV).  The strong reduction of the NCP over this small field range is a consequence of the sensitivity of the nuclei to fields comparable to that of the Knight field.


This rather surprising effect results from a nuclear field, ${\bf B}_N$, on the $e$ that effectively multiplies the effect of the external field by several orders of magnitude, and may be explained as follows.  When an external Voigt field is applied, the nuclei experience the
vector addition of the Knight and the external fields, ${\bf b}_T={\bf
b}_e + {\bf b}_x$, as shown schematically in the inset of Fig.~\ref{Fig2}(b). The nuclear field, ${\bf B}_N$, is  parallel to ${\bf b}_T$.  The strength of ${\bf B}_N$ is several orders of magnitude larger than field ${\bf b}_T$. As a consequence, optically oriented $e$'s precess around this field in a similar way to the behavior observed in Fig.~\ref{Fig2}(a), so that
depolarization occurs.  For $b_x >> b_e$, which is valid in our case for $b_x > 1$ mT, the precession direction will be very close to the $x$-axis.  As one can see in Fig.~\ref{Fig2}(b), this NCP
depolarization thus becomes saturated at $b_x > 1$~mT.  A similar effect has been observed for $e$'s localised on donors in GaAs \cite{Paget77}.

We therefore observe that, while at $b_x =$~0 T the co-aligned $e$-nuclear system leads to an increase in NCP, a strongly obliquely aligned nuclear field has a detrimental effect, causing a decrease in NCP to a value that is actually lower than if the nuclei were not polarized.
 Fig.~\ref{Fig2}(c) demonstrates this: a constant Voigt geometry field of ${\bf b}_x = 0.6$~mT is applied, resulting in a strongly obliquely aligned nuclear field (see the value at 0.6 mT in Fig.~\ref{Fig2}(b)).  A Faraday field $b_z$ is then applied, and for values of $b_z > b_x$ the polarization is restored to $~ -32\%$.  From the experimental data at high values of $b_z$, the expected NCP is extrapolated to zero $b_z$ field (dotted line): a value of $\sim -24\%$ is expected at $b_z =$ 0 T.  The experimental data deviates strongly from this behavior however, and, moreover, exhibits an asymmetry that is reversed upon reversal of the excitation helicity.
 
This asymmetry results from the partial cancellation of the Knight field, and therefore depolarization of the nuclei.  At a certain point, the Faraday field ${\bf b}_z$ is equal and opposite to the Knight field ${\bf b}_e$.  The Knight field is necessary for overcoming the dipole-dipole interaction, and thus when it is cancelled the nuclei quickly depolarize.  The result is in fact an increase in NCP as the nuclei depolarize: without the strong oblique field from the nuclei the $e$'s retain a greater polarization.  Verification of this is gained by the reversal in asymmetry when exciting with $\sigma^+$ (solid black circles) or $\sigma^-$ (open grey circles): reversal of helicity results in reversal of the Knight field.  Note that a fairly broad resonance is observed: as each nucleus in the QD feels a different value of Knight field, the entire nuclear system does not become simultaneously depolarized.  An estimate of the Knight field strength of $\sim 1$~mT may however be gained from the data, which agrees well with that observed in the similar system of $e$'s localized on donors in GaAs \cite{Paget77}.

\begin{figure}
\includegraphics[width=8cm]{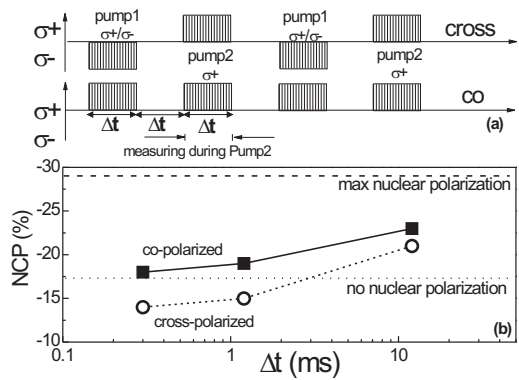}%
\caption{(a) Excitation schematic for co-polarized (lower) and cross-polarized (upper) Pumps 1 and 2.  NCP is measured during Pump 2 only. (b) NCP vs modulation time $\Delta t$ for Pumps 1 and 2
co-polarized (squares) and cross-polarized (circles). Dotted line
indicates NCP achieved for modulation frequency 50 kHz and
cross-polarized Pumps 1 and 2. Dashed line indicates value achieved
for no excitation modulation and co-circular polarization.}
\label{Fig3}
\end{figure}

We observe from the results in Fig.~\ref{Fig2} that the $e$ and nuclear spin systems are highly co-dependent.  I.~A.~Merkulov \cite{Merkulov98} makes the suggestion that in a strongly cooled nuclear system (a nuclear polaron) a giant increase of the nuclear spin relaxation lifetime should be observed.  As the $e$ polarization is strongly dependent on the feedback from the nuclei, its spin lifetime may also be increased. To investigate these
spin lifetimes, we modulate the excitation over time scales for the hyperfine interaction to act
efficiently. The sample is excited by trains of pulses (with a 13
ns pulse separation) which have duration from 20 $\mu$s to 12 ms. The
modulation is done by either a photoelastic modulator at 50 kHz or
by a mechanical chopper at $10-10^3$~Hz. Fig.~\ref{Fig3} (a) shows
the illumination scheme: the sample is excited for a time $\Delta
t$ by a first train of pulses (pump 1), and then by a second train
(pump 2), with a dark period between the trains of the same
duration, $\Delta t$.

Pumps 1 and 2 are in addition delayed by 3~ns with respect to each
other during the 13~ns laser repetition rate, allowing the PL from
each to be separated using a streak camera.  NCP was analysed from
the time-resolved PL spectra after the initial PL risetime.  The helicity of Pump 2 was kept at $\sigma +$ and its
NCP monitored as the polarization of Pump 1 is changed from
$\sigma +$ (co-polarized) to $\sigma -$ (cross-polarized).  Due to
the long $\Delta t$, each pump train contains between $10^3$ and
$10^6$ individual pulses. Without nuclear effects, we presume that
the resident $e$ spin polarization in the ensemble reaches a
steady state value well before the end of the pulse train
(remember NCP saturation as function of pump power). However, due
to the hyperfine interaction induced nuclear polarization
occurring over timescale of $10^{-2}$~s, influencing the
back-coupling to the $e$, we expect a strong dependence of NCP on
illumination time $\Delta t$.  For co-polarized excitation, both
pumps reinforce each others' orientation to allow maximum
polarization, whereas in the cross-polarized case the two pumps
act in competition.

Fig.~\ref{Fig3}(b) shows the NCP values as function of modulation
time $\Delta t$.  Let us first consider two limiting cases: The
dotted line indicates the NCP value ($\sim -17\%$) obtained when
Pumps 1 and 2 are cross-polarized, and modulated at 50 kHz. During
this $\Delta t$ ($\sim 1000$ pulses) the nuclei experience a field
${\bf b}_e$ that fluctuates over the timescale of $20 ~\mu$s, so
that the nuclei should not become polarized. Instead, the $e$'s
experience a randomly polarized nuclear field of $\sim 30$~mT
\cite{OultonNANO}. In contrast, the dashed line indicates the NCP
achieved without dark time and co-circular polarization,
exploiting the hyperfine interaction to its full degree. This value gives
the maximum NCP achievable of $-28\%$ at $B$ = 0~T. Thus, the $e$
polarization is almost doubled by the nuclear polarization.

Let us now consider the NCP achieved for intermediate modulation
times $\Delta t$. The squares in Fig.~\ref{Fig3}(b) show the NCP for
co-polarized pumps. The polarization increases from $\sim -17\%$,
to $\sim -23\%$. As $\Delta t$ ranges from below to above the time
scale for hyperfine effects ($10^{-2}$ s), this NCP increase is expected. At the
beginning of illumination the spin-polarized $e$'s begin to polarize
the nuclei through flip-flop processes. The longer the
illumination lasts, the more nuclear polarization will accumulate,
which in turn stabilizes the $e$ polarization, determining also
the NCP.

During the dark time, however, some nuclear spin polarization is
lost. If full nuclear polarization were retained during
the dark time, it would accumulate over the pump pulse cycles in the co-polarized case, and reach the unmodulated value of $\sim -28\%$. If
on the other hand no polarization memory remained from the
last pump, the NCP achieved would be identical for both co and cross pumping
protocols. That some nuclear polarization is maintained can be
seen in Fig.~\ref{Fig3}(b) from the comparison of the data for co- and
cross-polarized excitation. A difference
in NCP is evident even for $\Delta t$ = 20 ms. From the decrease
of the difference in NCP for both protocols a spin
memory time in the ms-range may be evaluated.

To investigate this long spin memory, let us consider the spin dynamics during the dark time. If
there were no nuclear polarization, the $e$ in each dot would
precess about the random nuclear field, which fluctuates on a
$\mu$s scale: on a ms timescale all $e$ polarization would be lost. Due to the presence of finite nuclear polarization along
the optical propagation direction, the influence of the random
field is reduced, and the $e$ spin is stabilized. This is in
turn important for the nuclei: an $e$ spin depolarization
would immediately reduce the nuclear polarization, as has been
demonstrated in Fig.~\ref{Fig2}. The spin lifetime lies therefore for both
$e$ and nuclei in the ms-range due to mutual stabilization.
This is orders of magnitude longer than for either decoupled
system.

Under conditions of a cooled nuclear spin system, spin
fluctuations of both nuclei and $e$ should be greatly reduced.
Theoretical estimates \cite{Merkulov98} suggest that after nuclear
cooling by optical pumping, a nuclear spin polaron may be formed for localized
$e$'s at crystal temperatures of 5~K and $e$ spin polarization $>
20\%$, close to the situation in our experiment. We suggest
therefore, that a nuclear polaron is formed between the $e$ and
the nuclei in the QD due to their strong interaction, which is,
however quite fragile: A small external perturbation to this
system results in strong depolarization of the $e$ and nuclei. It appears therefore that the predicted "giant increase of the nuclear spin relaxation time" suggested in Ref. \cite{Merkulov98} is responsible for the millisecond NCP lifetimes measured, and is therefore an extremely good indication that a nuclear spin polaron is formed in this system.

This work was supported by the BMBF, DFG and DARPA. R.O.
thanks the Alexander von Humboldt foundation.
We also thank I.A. Merkulov for discussions.

\end{document}